\documentclass[%
 reprint,
 amsmath,amssymb,
 aps,
prapplied,
floatfix,
]{revtex4-2}

\usepackage{graphicx}
\usepackage{dcolumn}
\usepackage{bm}
\usepackage{url}
\usepackage{xcolor}
\usepackage[export]{adjustbox}

\usepackage{hyperref}

\usepackage{lineno}

\begin{document}

\title{Tunable responsivity and bandwidth in microwave kinetic inductance detectors via readout current nonlinearity}

\author{M. Rouble}
\affiliation{t0.technology, Montreal, QC, H2X 0A4, Canada}
\email{maclean.rouble@t0.technology}

\author{P. Day}
\affiliation{Jet Propulsion Laboratory, California Institute of Technology, 4800 Oak Grove Drive,
Pasadena, California 91109, USA}

\author{M. Dobbs}
\affiliation{Department of Physics and Trottier Space Institute, McGill University, Montreal, QC, H3A 2T8, Canada}

\author{J. Montgomery}
\affiliation{t0.technology, Montreal, QC, H2X 0A4, Canada}

\author{S. Savchyn}
\affiliation{t0.technology, Montreal, QC, H2X 0A4, Canada}

\author{G. Smecher}
\affiliation{t0.technology, Montreal, QC, H2X 0A4, Canada}



\date{\today}

\begin{abstract}
{Microwave Kinetic Inductance Detectors (MKIDs) are generally read out with {microwave readout} tones of high enough amplitude to adequately {raise the signal above the} noise contribution of the {first-stage} amplifier.} 
{At high {readout} power, the detector's resonan{t} frequency is altered as a result of the dependence of the kinetic inductance on {the} internally circulating microwave current.  With the tone placed below the resonance frequency, the nonlinear frequency shift results in a positive feedback effect {that} can significantly enhance the respons{ivity} of the detector{, to both optical and microwave power}.  We report {an order of magnitude} enhancement in optical response by tuning the {readout} power and frequency to close to the resonator's bifurcation point. A corresponding decrease in the bandwidth of the resonator {is} observed under these conditions.  We show that the strength of the feedback effect can be easily selected by adjusting the excitation{, and provide a map of possible operational states}. Operation of MKIDs in this mode could be used to improve sensitivity when non-intrinsic noise sources are significant.}

\end{abstract}

\maketitle

\section{Introduction}\label{sec:introduction}

Superconducting resonator detectors such as microwave kinetic inductance detectors (MKIDs) are attractive for modern astronomical instruments and other cryogenic sensing experiments due to their high sensitivity, comparative ease of fabrication, and natural frequency-domain multiplexability.\cite{day2003}\cite{zmuidzinas2012}\cite{Baselmans2012KineticInductanceDetectors} They are seeing increasing adoption in power-integrating astronomical observations such as millimetre line intensity mapping as well as single-photon, energy-resolving or event-counting applications, such as astronomical instruments at UVOIR wavelengths or dark matter searches.\cite{doyle2008_lekids}\cite{mazin2019_review_MKIDs_2020s}\cite{day2024_25um_singlephoton} 
MKID arrays consist of large numbers of superconducting resonator-based detectors that are coupled to a common feedline and read out simultaneously. The maximum number of detectors which may be operated per feedline is presently limited in part by the available digitizable readout bandwidth and fabrication non-uniformity that leads to irregular frequency placement and inter-resonator interference.\cite{shu_trimming_2018}\cite{vissers2024}

Limitations to the sensitivity of an MKID include noise contributions from the first-stage amplifier and from fluctuations in the resonator's component materials that produce low-frequency variations in the resonant frequency.  Increased readout tone (hereafter the \textit{generator}) amplitude reduces the impact of additive noise sources such as the amplifier on the resonator phase determination.  Increased generator power also saturates the dielectric-related frequency noise in a way that is consistent with the two-level-systems theory of amorphous dielectrics. \cite{Gao2007NoisePropertiesCPWResonators} There is thus an incentive to operate with as much readout power as the detector will tolerate.

On the other hand, the readout power also modifies the resonator by altering its kinetic inductance. To leading order, this readout current nonlinearity can be described as a quadratic correction to the kinetic inductance, such that the induced change satisfies $\delta L_k \propto I^2$.\cite{pippard1950} The resulting resonant frequency shift, distorts the resonance lineshape measured with a high-power probe.  Past a critical power $P_c$ bistable hysteretic behaviour referred to as resonance bifurcation is observed.\cite{Swenson2013}\cite{deVisser2010} 
The onset of these effects is typically treated as an upper bound on the readout power that may be applied to a detector.

The nonlinear response of superconducting resonators has been extensively studied in the context of frequency-domain characterization.\cite{goldie2013}\cite{thomas2020_nonlinear_effects}\cite{Duell2024CCATNonlinearEffects}\cite{Valenti2019InterplayKineticInductance} 
It has been shown that the behaviour of a resonator operated deep into the nonlinear regime is well-modeled by a current-dependent kinetic inductance (with this current dependence arising from quasiparticle excitation, the inherent superconductor quadratic nonlinearity, or both), and improvements in detector performance have been reported when operating in this regime.
Here we take a complementary view of the nonlinearity and treat the alterations to the driven resonator response as a tunable operating resource, and derive a parametrization of the driven resonator state that is more suited for nonlinear detector operation than swept-frequency characterization.

In typical MKID operation, each readout tone is placed at a frequency that remains fixed on timescales significantly longer than the detector time constants. 
The parameter space spanned by the generator power and its frequency offset from the undriven resonant frequency determines the stored energy in the resonator and thereby its current-induced resonant frequency shift. In this basis,
the system exhibits positive or negative feedback between the driven resonant frequency and the readout current, depending on the position of the driven resonant frequency with respect to the readout tone. When the driven resonant frequency is above the tone, an increase in optical power drives it toward the tone, lowering the impedance at the readout frequency and increasing the current through the resonator. This in turn increases the shift in the resonant frequency. This is a positive feedback mechanism. When the driven resonance is below the readout tone, the converse is true: increasing optical load moves the resonance further from the tone, reducing the nonlinear contribution to the kinetic inductance.

Under positive feedback conditions, responsivity is enhanced and the dynamical response of the resonator is altered, extending the effective resonator time constant with observable consequences for detector bandwidth and its steady-state noise spectrum. Unlike in the conventional application of large readout amplitudes, the improvements in sensitivity attained in this way are not due solely to multiplicative scaling of the signal amplitude above system noise sources, but instead arise from alterations in the underlying physics of the driven resonator. This motivates {optimizing} the nonlinear bias point, rather than {treating it} as a secondary consequence of using high readout power.

This perspective is also relevant for improving the multiplexability of large arrays. High multiplexing factors are a central advantage of MKIDs, but high-density arrays are vulnerable to inter-resonator interactions. These arise both due to fabrication scatter and dynamic effects under changing optical loads. Recent work has shown that readout current can be used to deliberately alter and maintain the operating points of these detectors,\cite{rouble2024activefeedback} which if deployed at scale, would {reduce} dynamic resonator collisions and enable reliable expansion {of the multiplexing factor}. {In the sections that follow, we} develop an understanding of the principles underpinning the choice of operating point, and the consequences of that operating point on the detector's responsivity and bandwidth{, so that these techniques can be used effectively.}

\section{Driven nonlinear operating states}\label{sec:linear_and_nonlinear_response}

In this section we describe the driven resonator state using two experimentally controlled quantities: the detuning $x_0$ of the generator from the relaxed resonant frequency, and the generator power, $P_g$. Because of the current-dependent inductance, the driven resonant frequency depends on the stored energy in the resonator. Therefore, when a perturbation to the underlying resonator (e.g. through the quasiparticle population) changes its relaxed resonant frequency, this alters the readout current flowing through the inductor, which in turn alters the resulting driven resonant frequency.
We explore the modifications to the driven resonator's steady state due to this feedback, and its effects on the system's small-signal responsivity and bandwidth.

\subsection{Measurement setup}\label{sec:measurement_setup}

\begin{figure}[htbp]
    \centering
    \includegraphics[width=\linewidth]{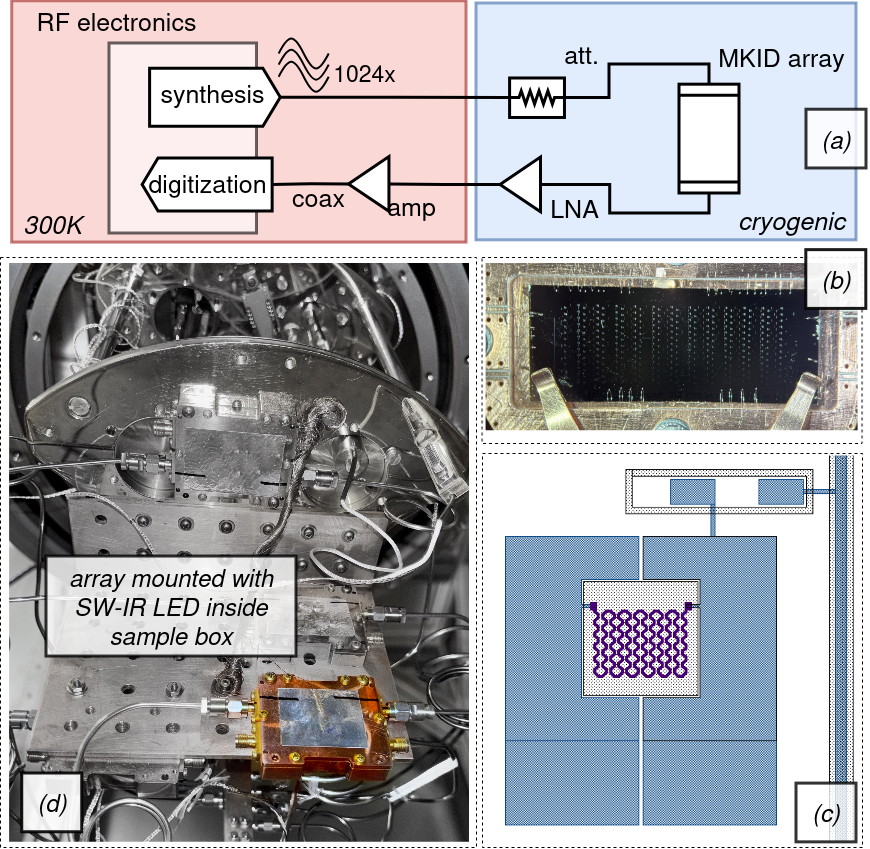}
    \caption{{Experimental configuration used in the study of driven nonlinear resonator operating states. (a) Schematic overview of the microwave readout chain: RF synthesis and digitization of generator tones is provided by a Control \& Readout System (CRS) board.\cite{Montgomery2024CRS} Attenuation and amplification stages condition the tones before and after the MKID array, respectively. The array is cooled to 100mK within a BlueFors LD250S dilution refrigerator. The frequency and power of each individual generator tone define the experimental control coordinates used throughout this work. (b) Photograph of a mounted MKID array. A variety of resonator geometries and materials were studied for this work, with consistent results. All measurements presented in this paper are made using the pictured array. (c) Rendering of an example resonator from the pictured array. The design is a lumped-element resonator with parallel plate capacitors (Nb, NbN) and hairpin inductor (TiN). (d) The array mounted within the cryogenic test environment. The array is housed in a sealed, high-purity copper box. A small, short wave IR LED is installed inside the box lid, to provide a calibration stimulus for the detectors.}
    }
    \label{fig:measurement_setup}
\end{figure}

{The measurements described in this work follow a typical experimental configuration for frequency-domain multiplexed MKID readout, pictured in Fig. \ref{fig:measurement_setup}a. A Control \& Readout System (CRS) board\cite{Montgomery2024CRS} synthesizes a comb of generator tones at GHz frequencies. The tones are transmitted to the MKID array via coaxial cables, through several stages of attenuation. After the array, the tones are amplified by a cryogenic low-noise amplifier (LNA) and two stages of warm amplification.}

{The operating state of an individual resonator is prepared by selecting the detuning, $x_0$, of one of these generator tones from the relaxed resonant frequency of a target resonator, and then {varying} the power delivered by that tone, $P_g$, until the intended state is reached. These two quantities form the experimental control coordinates used throughout this work.}

{A number of arrays and test devices were studied during the preparation of this work, spanning a variety of resonator designs and using either aluminum or titanium nitride (TiN) as the active metal. Results were consistent across the devices tested. For clarity, all experimental results presented within this work are measurements of a single test array. This array consists of approximately 300 hybrid-metal lumped-element MKID resonators, with TiN hairpin inductors and niobium-nitride-capped niobium parallel plate capacitors (Fig. \ref{fig:measurement_setup}c). }

\subsection{Steady-state nonlinear response}\label{sec:nonlinear_cavity_response}

As a function of the frequency offset between the generator and the resonant frequency, $x_0 = \frac{\omega_g - \omega_{0}}{\omega_{0}}$, the familiar expression for the forward transmission of a side-coupled resonator is \cite{zmuidzinas2012}

\begin{equation}
    S_{21} = 1 - \dfrac{Q_r}{Q_c} \dfrac{1}{1 + 2j Q_r x_0} \quad{.}
\end{equation}

\noindent {where the total resonator quality factor $Q_r = [Q_i^{-1} + Q_c^{-1}]^{-1}$, with $Q_i$ the internal dissipation quality factor of the resonator and $Q_c$ the quality factor describing its coupling to the feedline.}

The absorption of pair breaking energy alters $\omega_0$ and to a lesser extent $Q_r$ through the conversion of Cooper pairs into quasiparticles, altering the impedance of the superconductor. In the devices discussed in this work, the alteration {due to changing optical load} {primarily} affects the kinetic inductance, and it is this reactive component that will be the focus of the analysis.

In addition to its dependence on absorbed optical load, the kinetic inductance has a higher-order dependence on the readout current flowing through the resonator:

\begin{equation}\label{eq:LK_Isquared}
    L_k(I) \sim L_k(I=0) \left[1 + \frac{I^2}{I_*^2} \right]
\end{equation} 

\noindent with $L_k(I=0)$ the zero-current value and $I_* = \sqrt{\frac{\pi N_0 \Delta_0^3}{\rho_N \hbar}} w t$ setting the scale of the current-dependent nonlinearity through the geometry of the inductor (width $w$ and thickness $t$) and the properties of the material (gap energy $\Delta_0$, single-spin density of states $N_0$, and normal-state resistivity $\rho_N$).\cite{zmuidzinas2012}
{While dissipative effects due to readout current have been well-documented in the literature,\cite{deVisser2010}\cite{goldie2013}\cite{thomas2020_nonlinear_effects} for the resonators studied for this work, we observed the response to be primarily reactive. This is also consistent with studies in the literature.\cite{Swenson2013}\cite{Anferov2020}. For simplicity, we therefore consider only the reactive effects of the readout current in this analysis. To build a more complete model, future work should seek to include the dissipative components as well.}

When the current through the resonator is non-negligible, therefore, the nonlinear frequency shift is accounted for by modifying the expression for the resonator transmission as

\begin{equation}\label{eq:s21_nonlinear}
    S_{21} = 1 - \dfrac{Q_r}{Q_c} \dfrac{1}{1 + 2j Q_r x} \quad{,}
\end{equation}

\noindent where $x = \dfrac{\omega_g - \omega_r}{\omega_r}$ is the detuning of the generator tone from the shifted resonant frequency, $\omega_r = \omega_0 + \delta \omega_r$. To first order, we then have $x \simeq x_0 - \delta x$.
The nonlinear resonant frequency shift, $\delta x$ is proportional to the energy stored in the resonator, $E$. So:

\begin{equation}\label{eq:x_EoverEstar}
    x \simeq x_0 + \dfrac{E}{E_*}
\end{equation}

\noindent with $E_* \propto \left(\frac{I_*}{\alpha_k} \right)^2 L_k(I=0) $.

The Hamiltonian of a nonlinear driven resonant cavity contains terms of the form $\frac{\hbar}{2} K (a^\dagger a)^2$, which shift the resonant frequency by the cavity Kerr coefficient $K$ for every photon added to the cavity: $\omega_r(n_{ph}) = \omega_0 + K n_{ph}$.\cite{Anferov2020}\cite{YurkeBuks2006} 
Then, to first order, the nonlinear detuning is:

\begin{equation}\label{eq:x_Knph_over_omega0}
    x \simeq x_0 - \dfrac{Kn_{ph}}{\omega_{0}} \quad{.}
\end{equation}

\begin{figure}[htbp]
    \centering
    \includegraphics[width=0.5\linewidth]{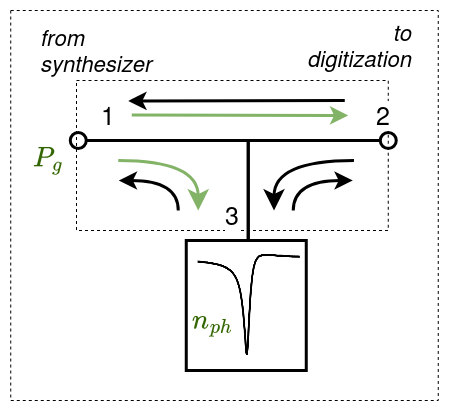}
    \caption{Schematic representation of the side-coupled resonator model. The generator tone enters port 1 with power $P_g$, the transmitted signal ($S_{21}$) is measured at port 2, and port 3 couples the generator to the resonator mode with photon occupation $n_{ph}$. The three-port network maps the feedline-coupled resonator to the one-port Kerr-cavity input-output model used to describe the nonlinear state. {The arrows illustrate the three-port scattering matrix, with the ones relevant for this discussion highlighted in green.}}
    \label{fig:cavity_model}
\end{figure}

As described in \cite{Anferov2020}, the resonator may be understood as the one-port optical cavity,\cite{YurkeBuks2006}\cite{Eichler2014} here coupled symmetrically into a three-port network as shown in Fig. \ref{fig:cavity_model}. 

{F}rom  Eqs. \ref{eq:x_EoverEstar} and \ref{eq:x_Knph_over_omega0}, $\delta x \simeq \frac{Kn_{ph}}{\omega_0}$, with $K=\frac{-\hbar \omega_0^2}{E_*}$ and $n_{ph} = \frac{E}{\hbar \omega_0}$, the average photon occupancy of the cavity. 
Using the standard input-output equation for the side-coupled geometry in Fig. \ref{fig:cavity_model} (\cite{Eichler2014}; Appendix \ref{app:steady_state_derivation}), the steady state photon number satisfies:

\begin{equation}\label{eq:steady_state_nph_Pg}
    n_{ph} \hbar \omega_g \dfrac{2 Q_c}{\omega_0}  \left[ (\omega_g - \omega_0 - Kn_{ph})^2 + \left(\dfrac{\omega_0}{2 Q_r}\right)^2  \right] = P_g \quad{.}
\end{equation}

\begin{figure}[htbp]
    \centering
    \includegraphics[width=\linewidth]{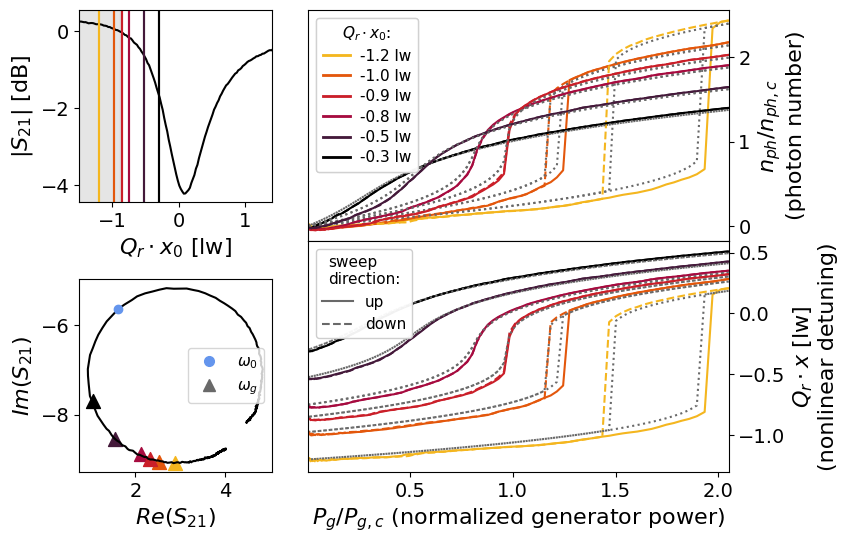}
    \caption{{Measured response of a resonator to generator power sweeps, performed} at a selection of generator frequency offsets,  $x_0$ (here scaled by $Q_r$ to express the offset in units of linewidths), from the undriven resonance. The left panels illustrate the setup of each power sweep, overlaid on a low-power $S_{21}$ frequency sweep {of the same resonator}. The right column shows the response of the resonator to the power sweep when the readout tone is fixed at each value of $x_0$, in terms of the average number of photons in the resonant cavity $n_{ph}$ (top) and nonlinear detuning $x$ (bottom). For values of $x_0$ where the generator is located close to the undriven resonant frequency, the response is smooth and single-valued, with both an ascending {(solid lines)} and descending {(dashed lines)} power sweep tracing the same values. For larger values of $x_0$ (the generator located further below the undriven resonant frequency{, at or beyond the critical point, indicated by a grey shaded region}), the response bifurcates and shows a hysteresis depending on the direction of the sweep. {Grey dotted lines are solutions to Eqs. \ref{eq:s21_nonlinear} and \ref{eq:steady_state_nph_Pg} for the measured resonator.}}
    \label{fig:nph_vs_Pg}
\end{figure}

In the following discussions, we use $n_{ph}$ primarily as the Kerr cavity state variable, as it defines the nonlinear frequency shift $Kn_{ph}$ and is the natural coordinate in which to express the steady state response, the conditions for bifurcation, and the dynamics of the linearized system. To apply these principles to detector operation, however, it is more convenient to work in terms of the nonlinear detuning $x$ which results from the Kerr frequency shift, as $x$ describes the position of the readout tone within the resonance bandwidth. The response of the resonator to applied readout power in terms of both $x$ and $n_{ph}$ is shown in Fig. \ref{fig:nph_vs_Pg}.

For a generator fixed at $x_0$, sweeping $P_g$ moves the resonance with respect to the generator as the number of photons in the cavity increases or decreases. This may equivalently be understood in terms of the nonlinear kinetic inductance increasing or decreasing, depending on the current through the cavity. For small $|x_0|$, the response is monotonic and a given $(x_0, P_g)$ selects a unique driven state. For sufficiently negative $x_0$, the response steepens until there is a region where $\frac{dn_{ph}}{dP_g} \rightarrow \infty$. For $x_0$ beyond this point, the response is hysteretic and no longer single-valued: the same applied generator power corresponds to more than one stable driven state. This is the same bifurcation behaviour which is familiar from swept-frequency measurements.\cite{Swenson2013}\cite{deVisser2010}

\subsection{Power sweep steepening and the critical point}\label{sec:feedback_and_the_critical_point}

For a generator fixed at $x_0$, increasing $P_g$ increases the stored energy in the resonator: $\frac{d n_{ph}}{d P_g} > 0$. As the system approaches the onset of bifurcation, this derivative becomes large, with a small change in the applied readout power producing a large change in the stored energy, and thus a large resulting frequency shift.
This point is a cusp or critical point in the readout power --- nonlinear detuning parameter space. From Eq. \ref{eq:steady_state_nph_Pg}, the coordinates of this cusp are easily solved for using the inverse response, $P_g(n_{ph};x_0)$, at fixed drive frequency. At the critical point,

\begin{equation}
    \dfrac{d P_g}{dn_{ph}} \rightarrow 0 \quad \mathrm{and} \quad  \dfrac{d^2P_g}{dn_{ph}^2} \rightarrow 0 \quad{.}
\end{equation}

The critical frequency offset between the generator and the relaxed resonant frequency is:

\begin{equation}\label{eq:foffset_crit}
    x_{0,c} = \dfrac{\sqrt{3}}{2 Q_r} \dfrac{K}{|K|}
\end{equation}

\noindent and, at this generator frequency, the cusp is reached when the nonlinear shifted resonant frequency is detuned from the generator by:

\begin{equation}
    x_c = \dfrac{1}{2 \sqrt{3} Q_r} \dfrac{K}{|K|} \quad{.}
\end{equation}

The critical number of photons is:

\begin{equation}
    n_{ph, c} = \dfrac{1}{\sqrt{3} |K|} \dfrac{\omega_0}{Q_r}
\end{equation}

\noindent and the critical generator power is:

\begin{equation}
    P_{g,c} = \dfrac{2 \hbar \omega_g}{3\sqrt{3} |K| } \dfrac{Q_c \omega_0^2}{Q_r^3} \quad{.}
\end{equation}

$x_{0}$ thus defines the region where bistable behaviour may occur. For offsets beyond $x_{0,c}$, there is a range of generator powers for which two stable driven states exist. {This is seen in Fig. \ref{fig:nph_vs_Pg} on traces where the ascending and descending power sweeps are not identical; $x_0$ in these cases is beyond the critical value.}
For generator offsets within the critical value, the response of the driven resonator to a power sweep is always single-valued, although may still exhibit strong nonlinearity.

\subsection{Small-signal responsivity}\label{sec:small_signal_responsivity}

The energy stored in the resonator depends on the generator power and frequency, {operating conditions such as temperature and incident light, and} the properties of the resonator {itself}:\cite{Swenson2013} 

\begin{equation}\label{eq:E}
    E = \dfrac{2 Q_r^2}{Q_c} \dfrac{1}{1 + 4 Q_r^2 x^2}\dfrac{P_g}{\omega_r}
\end{equation}

\noindent where {$Q_r$ depends on $Q_i$,} the internal dissipation quality factor of the resonator{, which is affected by absorbed optical and thermal energy}.
{Letting $\Tilde{E} \equiv \frac{E}{E_*}$,} we may rewrite Eq. \ref{eq:x_EoverEstar} as:

\begin{equation}
    x = x_0 + \Tilde{E}(x, P_g, Q_i^{-1})
\end{equation}

A small fluctuation produces:

\begin{equation}\label{eq:deltax_all}
    \delta x = \dfrac{\delta x_0 + \frac{\partial \Tilde{E}}{\partial P_g} \delta P_g + \frac{\partial \Tilde{E}}{\partial Q_i^{-1}} \delta Q_i^{-1}}{1 - \frac{\partial \Tilde{E}}{\partial x}}
\end{equation}

\noindent where $\delta x_0$ and $\delta Q_i^{-1}$ are the resulting changes in the undriven resonant frequency and internal dissipation respectively.
Swenson \cite{Swenson2013} also provides the derivatives:

\begin{equation}
    \dfrac{\partial \Tilde{E}}{\partial x} = \left[ \dfrac{1}{1+x} - \dfrac{8Q_r^2x}{1 + 4Q_r^2 x^2}  \right] \Tilde{E}
\end{equation}

\noindent and

\begin{equation}
    \dfrac{\partial \Tilde{E}}{\partial Q_i^{-1}} = \frac{-2Q_r}{1 + 4Q_r^2 x^2} \Tilde{E} \quad{.}
\end{equation}

{Eq. \ref{eq:deltax_all} contains several useful small-signal limits. The responsivity to optical load at fixed generator power is obtained by setting $\delta P_g = 0$, allowing the optical perturbation to change both $x_0$ and $Q_i^{-1}$.
Similarly, the responsivity to a change in generator power at fixed optical load is obtained by setting $\delta x_0 = \delta Q_i^{-1} = 0$.   }

{The detector's responsivity to the generator is distinct from the responsivity a signal fluctuation. The discussion in Sec. \ref{sec:feedback_and_the_critical_point} describes the system's response as the generator power is swept under fixed optical load. In {typical} MKID operation, the generator power and frequency are both fixed and absorbed optical load changes the quasiparticle population, producing changes in the relaxed resonant frequency $\omega_0$ and the internal quality factor of the resonator, $Q_i$. The measured operational response is then the resulting change in the total driven resonant frequency after the readout current has adjusted to the new state of the resonator.}
{The numerators of the two responsivity quantities reflect the different physical properties between them, but they share a common denominator, $1-\frac{\partial \Tilde{E}}{\partial x}$. This denominator is the reactive readout current feedback factor.}

{The feedback factor describes the modification in the stored energy due to a small change in detuning, which then feeds back into the nonlinear shifted resonant frequency. I}f the perturbation changes the detuning in a way which increases the stored energy, the denominator becomes small, and the resulting change in the nonlinear detuning is enhanced as the readout current change reinforces the perturbation. If it decreases the stored energy, the induced readout current change opposes the perturbation, suppressing the response.

Where {the feedback factor} {is less than unity}, the system's response to applied power from both the readout and optical load is enhanced. The derivative $\partial \Tilde{E} / \partial x$ may be considered to be the loop gain of the driven resonator system. When a perturbation changes the detuning $x$, the resulting change in the stored energy $E$ feeds back into $x$. For $1 > \partial \Tilde{E} / \partial x > 0$, the feedback is positive: the induced change in stored energy reinforces the original perturbation, enhancing the response. For $\partial \Tilde{E} / \partial x < 0$, the feedback is negative, and suppresses the response. As $\partial \Tilde{E} / \partial x \rightarrow 1$, the response becomes singular and the loop is unstable. This is the small-signal equivalent to the cusp in the power sweeps discussed in Sec. \ref{sec:feedback_and_the_critical_point}. In the multivalued region beyond the cusp, the unstable branch corresponds to $\partial \Tilde{E} / \partial x > 1$, while the two stable observable branches have perturbations which decay toward the chosen driven state.

The responsivity of the driven system to optical load therefore depends on both the undriven resonator's optical response and on the driven state. It is worth noting that the peak responsivity occurs where the feedback factor {$1 - \frac{\partial \Tilde{E}}{\partial x}$} approaches zero, and that this does not generally coincide with zero detuning.

Defining the low-power and driven responsivities of the detector as, respectively:

\begin{equation}
    R_{x,0} = \dfrac{\partial x_0}{\partial P_{opt}} \quad \mathrm{and} \quad R_x=\dfrac{\partial x}{\partial P_{opt}}
\end{equation}

\noindent then we can express {the constant readout power limit of Eq. \ref{eq:deltax_all}} as the ratio of the nonlinear responsivity to the undriven responsivity:

\begin{equation}\label{eq:nonlinear_optical_responsivity_normalized}
    \dfrac{R_x}{R_{x,0}} = \dfrac{1 + \frac{\partial \Tilde{E}}{\partial Q_i^{-1}} \frac{\partial Q_i^{-1}/\partial P_{opt}}{\partial x_0/\partial P_{opt}} }{1 - \frac{\partial \Tilde{E}}{\partial x}} \quad{.}
\end{equation}

\begin{figure}[htbp]
    \centering
    \includegraphics[width=\linewidth]{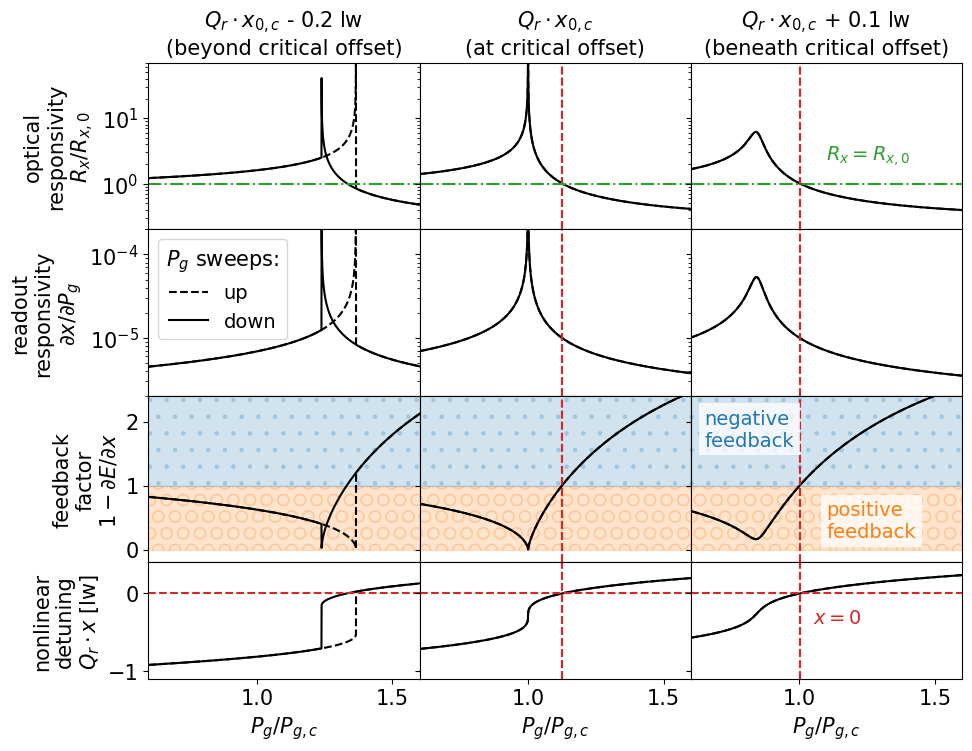}
    \caption{{Calculated} responsivity to optical load (Eq. \ref{eq:nonlinear_optical_responsivity_normalized}), readout ({constant-load limit of Eq. \ref{eq:deltax_all}}), {the feedback factor, and the nonlinear detuning, as a function of generator power,} for three values of $x_0$ beyond (left), at (centre), and beneath (right) the critical offset frequency. The optical responsivity (top row) and the readout responsivity (second row) are both maximized where the feedback factor (third row) is closest to zero. 
    {The feedback factor enhances the response} for $x<0$ {(orange dotted region)} and {suppresses it} for $x>0$ {(blue circles region)}; the optical responsivity is enhanced and suppressed in these regions respectively.
    {The ratio of this resonator's reactive to dissipative response is large, so} even for highly driven operating points at large $x_0$, the responsivity at $x=0$ (if it is accessible on the given branch) is equal to the undriven responsivity {(Eq. \ref{eq:responsivity_ratio_zero_detuning})}. 
    {Improved responsivity may be attained by instead choosing to operate at slightly negative detuning, at the cost of reduced bandwidth (Sec. \ref{sec:dynamics}).}
    }
    \label{fig:three_foffsets_responsivities}
\end{figure}

This ratio is shown in the upper row of Fig. \ref{fig:three_foffsets_responsivities} for fixed-frequency power sweeps with the generator located beyond, at, and within the critical $x_0$ (left, centre, right, respectively). This is compared with the readout responsivity ({fixed-load limit of Eq. \ref{eq:deltax_all}}) in the upper middle row. The lower middle row shows the reactive feedback factor, while the lowest row shows the nonlinear detuning at each corresponding point in the power sweep. Because of the common denominator, the readout and optical load responsivities share common features such as a prominent peak where the feedback factor approaches its minimum, but are not identical due to their distinct numerators.

{Typical MKID operation aligns the readout tone with the resonant frequency (i.e., at $x=0$), so it is worth paying extra attention to the behaviour at this point. Solving Eq. \ref{eq:nonlinear_optical_responsivity_normalized} at zero detuning, we obtain:}

\begin{equation}\label{eq:responsivity_ratio_zero_detuning}
    {\left. \dfrac{R_x}{R_{x,0}} \right|_{x=0} = \dfrac{1 -2 Q_r \Tilde{E} \beta^{-1}}{1 - \Tilde{E}}}
\end{equation}

\noindent {where $\beta \equiv \delta x_0 / \delta Q_i^{-1}$, the ratio of the undriven reactive and dissipative responsivities. Where $\beta$ is large, the ratio of the driven versus undriven responsivity at zero detuning remains close to unity. Resonators with a larger dissipative response (smaller $\beta$), however, will receive enhanced responsivity at this point. This effect increases with applied generator power, or equivalently, for zero detuning at larger $x_0$.}

{The value for $\beta$ used in the calculations in Fig. \ref{fig:three_foffsets_responsivities} is based on the measured response of resonators in the test array. Within the range of reasonable small-signal optical load variation that was applied to the array, a dissipative response was not discernible above measurement uncertainties. Thus we choose $\beta = 20$ as a placeholder value in the primarily-reactive regime.}

The discussion here pertains to small perturbations about a stable driven operating point, at frequencies sufficiently low that the resonator bandwidth is not relevant. However, as discussed in the following section, as the system is driven close to its critical point, the resonator dynamics are also affected by the nonlinear feedback and must be taken into consideration.

\subsection{Dynamics and critical slowing}\label{sec:dynamics}

The same feedback that enhances the static responsivity also changes the restoring rate of the driven resonator system. 
{Eq. \ref{eq:steady_state_nph_Pg} describes the steady state condition for the driven Kerr cavity resonator for a chosen $x_0$ and $P_g$. To examine the dynamics about this state, we consider a small perturbation to the complex resonator response. A full derivation from the current-dependent inductance is given in Appendix \ref{app:dynamics}. There, we express the system dynamics in terms of $U\equiv 1 - S_{21}$, which is proportional to the circulating current and field in the resonator. The steady state solution is $U_0$, which we perturb as $U(t) = U_0 + \epsilon(t)$, with $\epsilon(t) = \epsilon_x + j \epsilon_y$. To linear order, the perturbation obeys:}

\begin{equation}
    \dfrac{d}{dt} \begin{bmatrix}
        \epsilon_x \\
        \epsilon_y
    \end{bmatrix}
     = M \begin{bmatrix}
          \epsilon_x \\
        \epsilon_y
     \end{bmatrix}
\end{equation}

{where the Jacobian $M$ is evaluated at the chosen driven operating point. We decompose the perturbation into two eigenmodes of $M$, each of which evolves as $\vec{\epsilon} \propto e^{\lambda_{\pm}t} \vec{V}_{\pm}$.} 
{The eigenvalues $\lambda_{\pm}$ give the decay rates of the two modes, and the corresponding $\vec{V_{\pm}}$ give their directions in the complex $S_{21}$ plane of the resonator response. The eigenvalues are: }

\begin{equation}\label{eq:linearized_eigenvalues}
    \lambda_{\pm} = -\dfrac{\omega_0}{2 Q_r} \pm \sqrt{(Kn_{ph})^2 - (\omega_0 - \omega_g + 2Kn_{ph})^2} \quad{.}
\end{equation}

{Stability under a given fluctuation requires $Re(\lambda) < 0$, which leads to the decay of the perturbation mode. The time constant of the system for this mode is the inverse of the rate of this decay, $\tau_r = \frac{-1}{Re(\lambda)}$.}

In the low-power case, the square root term is imaginary, and so a perturbation {of either mode} in the system decays at $\tau_{r,0} = \frac{2Q_r}{\omega_0}$, the undriven resonator ring time.
{As the system is driven nearer to the critical point, the two decay rates separate, with one becoming progressively slower and the other remaining strongly damped. At the critical point derived in Sec. \ref{sec:feedback_and_the_critical_point},}

\begin{equation}
    \lambda_+ = 0 \quad \mathrm{and} \quad \lambda_- = \dfrac{-\omega_0}{Q_r} \quad{.}
\end{equation}

{The eigenvectors provide a geometric interpretation of this result. At the critical point, the slow eigenvector $\vec{V_+}$ is tangent to the resonance circle in the complex $S_{21}$ plane; that is, the frequency direction of the resonator's response. Perturbations in this direction therefore experience no first-order restoring force, and, in an ideal noiseless system, do not decay. Conversely, perturbations in the second eigenmode relax at twice the low-power rate.}

\begin{figure}
    \centering
    \includegraphics[width=\linewidth]{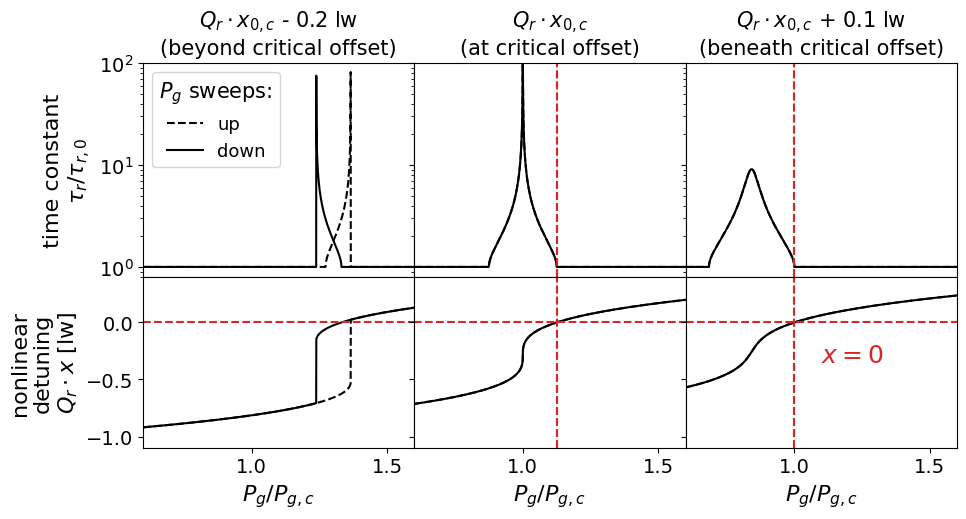}
    \caption{Driven resonator relaxation time during fixed-frequency power sweeps, for generator offsets beyond, at, and within the critical offset $x_{0,c}$. The relaxation time remains nearly constant and equal to the undriven resonator ring time, $\tau_{r,0}$, except near the critical point, where the slow eigenvalue becomes small and $\tau_r$ increases sharply. For $x_0$ beyond the critical value, the response is branch dependent because the power sweep is hysteretic.
    }
    \label{fig:time_constant_three_foffsets}
\end{figure}

{The relevant observable relaxation time is dominated by the slowest mode, and so near the critical point the response of the resonator decays at $\tau_r = \frac{-1}{Re(\lambda_+)}$. In an ideal system this approaches infinity as the critical point is attained. Elsewhere within the positive feedback regime, there is a range of generator powers (equivalently a range of nonlinear detunings) at which the time constant is substantially increased, as seen in Fig. \ref{fig:time_constant_three_foffsets}.}

Evaluating the time constant along a fixed frequency power sweep (Fig. \ref{fig:time_constant_three_foffsets}) shows a qualitatively similar structure as the optical responsivity. In the linear region, $\tau_r$ remains at or very close to the undriven ring time. As the operating point approaches the critical point, {the slow eigenvalue approaches zero and} $\tau_r$ rises sharply. For generator offsets beyond $x_{0,c}$, the time constant is branch-dependent, as the same generator power corresponds to two stable driven states. 

It should be noted that the $\tau_r$ discussed here is the driven resonator relaxation time, not the quasiparticle lifetime, $\tau_{qp}$. In the measured detector response, the spectrum of resonant frequency fluctuations resulting from quasiparticle fluctuations is filtered by both the dynamics of quasiparticle recombination and by the resonator ringtime. When $\tau_r \ll \tau_{qp}$, the usual quasiparticle lifetime sets the observed roll-off. Near the critical point, $\tau_r$ can become comparable to or longer than $\tau_{qp}$. This adds a lower frequency pole and slows the measured response. 

The increase in $\tau_r$ and the increase in low-frequency responsivity are not independent effects. Both arise because the same feedback denominator approaches zero near the cusp. For operational states in this region, the detector therefore trades bandwidth for low-frequency gain.

\subsection{Operating-state maps}\label{sec:operating_state_maps}

\begin{figure}[htbp]
    \centering
    \includegraphics[width=\linewidth]{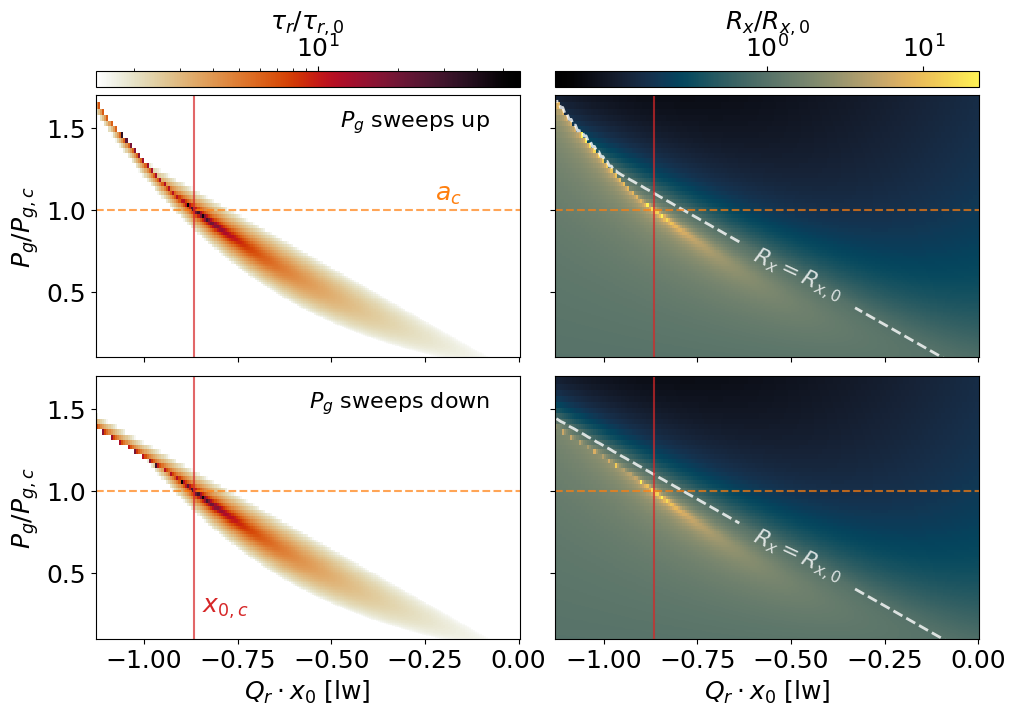}
    \caption{Properties of the nonlinear operating state in control-space coordinates: driven resonator relaxation time $\tau_r$ (left) and optical load responsivity $R_x$ (right), evaluated over a range of the experimentally-controlled parameters, $x_0$ and $P_g$. The generator power is normalized to the power at the critical point, and both plotted quantities are normalized to their undriven values. The upper maps show the stable branch traversed by an upward power sweep, with a downward sweep (lower panels) differing only in the hysteretic region beyond the critical point. A dashed line indicates the critical value of $a$, which is often used as the target for nonlinearity in conventional biasing procedures. In the region approaching the critical point, the time constant becomes large, and the optical load responsivity is enhanced.
  }
    \label{fig:taur_and_dfrdPopt_param_space}
\end{figure}

The experimentally controlled coordinates are the generator offset from the relaxed resonance, $x_0$, and its power, $P_g$. In this control-space view (Fig. \ref{fig:taur_and_dfrdPopt_param_space}) the region of increased relaxation time and responsivity appears as a narrow wedge which is peaked at the critical point. In Fig. \ref{fig:taur_and_dfrdPopt_param_space}, the generator power is swept upwards, although much of the plotted parameter space is single-valued. Sweeping down produces an identical plot except in the hysteretic regions beyond $x_{0,c}$ and $P_{g,c}$. 

\begin{figure}[htbp]
    \centering
    \includegraphics[width=\linewidth]{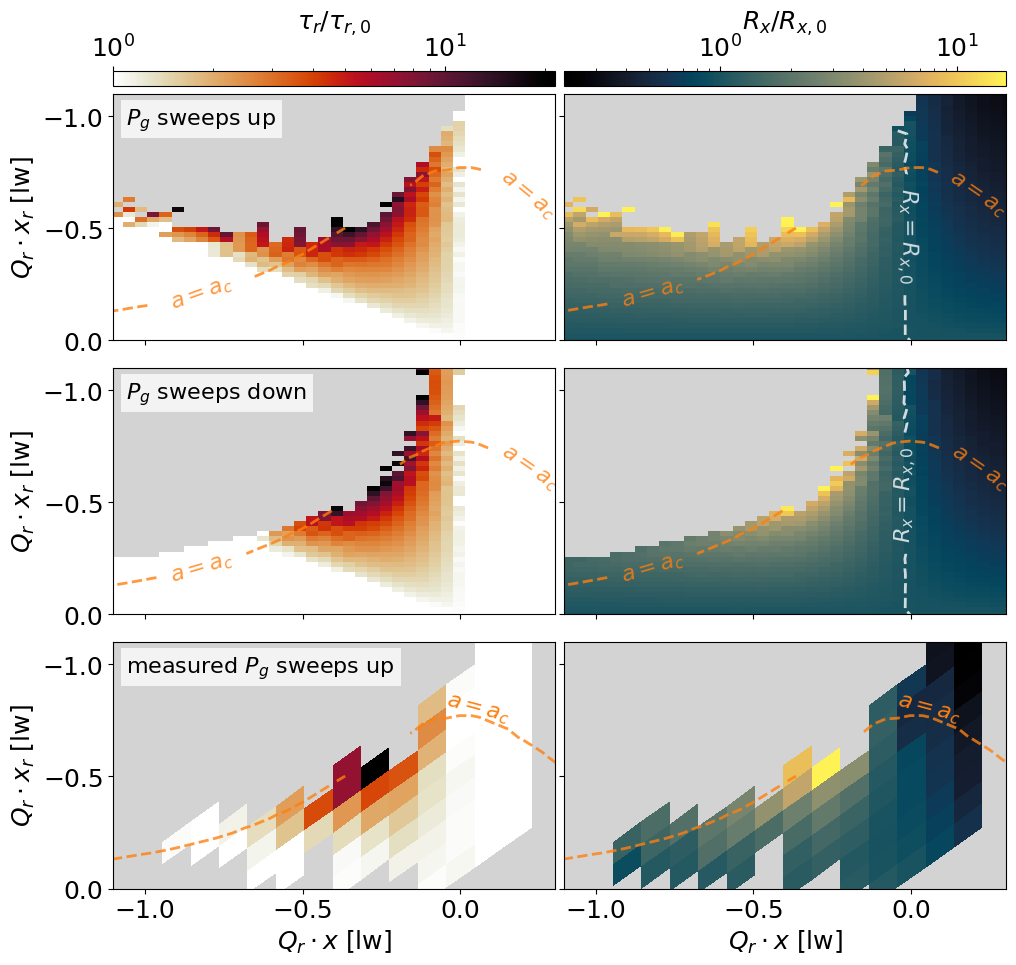}
    \caption{Driven-state coordinate view of the nonlinear resonator parameter space, a reparameterization of the $(x_0,P_g)$ control space map in Fig. \ref{fig:taur_and_dfrdPopt_param_space}. The resonator time constant $\tau_r$ (left column) and optical-load responsivity $R_x$ (right column), normalized to their low-power values, are shown as functions of the generator detuning from the driven resonance, $x$, and the nonlinear resonance shift, $x_r=(\omega_r-\omega_0)/\omega_0$, both in linewidths. The upper and middle rows show the stable branches reached during {calculated} upward and downward sweeps of generator power. {The lowest row shows measured data from a set of upward sweeps, performed simultaneously on a representative set of five resonators in the array. The data plotted here are the inverse-variance weighted mean across the detectors.}
    Grey regions indicate states not reached on the plotted branch{, or (in the case of the measured data) where the measurement uncertainty was large.}
    Orange dashed lines indicate the contour of $a=a_c=\frac{4\sqrt{3}}{9}$. Typical detector biasing aims to operate near the point $(x=0, a=a_c)$. At this point, the responsivity and time constant are unchanged from the undriven value. {\textbf{By instead selecting a bias point which is detuned from the driven resonant frequency, the responsivity is measured to increase by more than an order of magnitude, with a corresponding decrease in bandwidth. The largest measured mean responsivity increase for which at least 80\% of the detectors in the sub-array were operably stable was $R_x/R_{x,0} = 16 \pm 4$, associated with a time constant increase of $\tau_r/\tau_{r,0} = 17 \pm 4$.}}
}
    \label{fig:operating_space_tau_resp}
\end{figure}

The $(x_0, P_g)$ parameter space is convenient both for deriving the nonlinear steady state and for preparing the system to a chosen operating point.
However, to explore the driven resonator parameter space, it 
is useful to reframe the space in resonator-specific quantities (Fig. \ref{fig:operating_space_tau_resp}). Here, we consider the extent by which the resonator has been shifted by the nonlinearity as $x_r = \frac{\omega_r - \omega_0}{\omega_0}$, and, once this shift has been attained, the nonlinear detuning{, $x$,} between the driven resonant frequency and the readout tone. This latter parameter is the same nonlinear detuning as described previously, although the $x_0$ required to attain it at a given $x_r$ now varies as $x_0 \simeq x + x_r$.

This framing is conceptually consistent with the conventional KID biasing framework, in which the generator frequency is chosen relative to the resonant frequency based on the properties of the resonator transfer function as a function of frequency. The conventional, low-power case corresponds to $x_r = 0$ in Fig. \ref{fig:operating_space_tau_resp}. Typical high-power readout operates the system {at powers such that} {the asymmetry parameter} $a \rightarrow \frac{4\sqrt{3}}{9}$ \cite{Swenson2013} {(orange dashed contour) and zero detuning. Although the contour of $a \rightarrow a_c$ crosses a region of substantially increased responsivity, at $x=0$, the responsivity is unchanged from the undriven responsivity (for detectors with a predominately reactive response to optical load). Choosing to instead place the readout tone at slightly negative detuning would increase the detector responsivity, while reducing its bandwidth. } 

As can be seen {in both views of this parameter space, there is a region in which the resonator responsivity and time constant are significantly increased from their undriven values.} Thus, the improvements in sensitivity attained by operating there are not simply from multiplicatively scaling the signal amplitude by the carrier amplitude and increasing headroom above fixed LNA or system noise, but instead {from the alteration of the underlying resonator state itself, which produces a larger change in resonant frequency for a given input signal.}

\section{Resonator frequency noise spectrum}\label{sec:observables}

The analysis above predicts two experimental signals of positive feedback: an increase in the low-frequency response to quasiparticle fluctuations and a reduction in the bandwidth of that response. These are local properties of the driven operating point, and therefore can be probed using small fluctuations about the steady state, such as from quasiparticle number fluctuations. 

The spectrum of resonant frequency fluctuations due to the generation and recombination of thermalized quasiparticles is a commonly-used probe of detector properties.
The height of the white noise level of the spectrum reflects the scale of the resonant frequency shift resulting from a given quasiparticle number fluctuation, while the frequency at which this shelf rolls off represents the detector time constant. This roll-off is determined by the longer of the quasiparticle lifetime or the resonator ring time.

\begin{figure}[htbp]
    \centering
    \includegraphics[width=\linewidth]{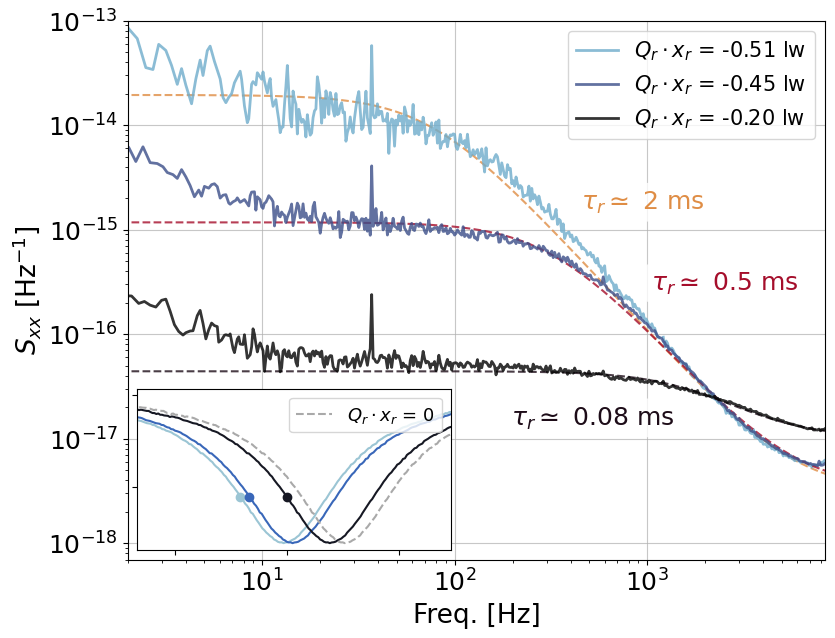}
    \caption{Spectrum of resonant frequency fluctuations under increasing degrees of nonlinear bias{, measured in the laboratory with a representative MKID from the test array}. {The inset illustrates the state of the resonance and readout tone providing the bias{: the low-power $S_{21}$ curve is shown shifted to the corresponding driven resonance location}. Three different tone locations are used at increasing offset from the relaxed resonant frequency ($x_{0,c} < x_0 < 0$). At each, $P_g$ is such that $x$ is constant between the spectra. This equates to an increase in the amount that the resonance has been shifted by the nonlinearity, $x_r$, i.e. moving vertically in Fig. \ref{fig:operating_space_tau_resp} at $Q_r x \simeq -0.4$ linewidths.
    As $x_r$ increases,} the resonator responsivity and time constant both increase, raising the white level of the measured spectrum and reducing the roll-off frequency, respectively. {A peak at 37 Hz is from a sinusoidally-varying cryogenic LED}; using this as a proxy for a small signal on the detector, we observe that the increase in responsivity applies equally to the intrinsic quasiparticle noise spectrum as to an extrinsic signal.
    }
    \label{fig:spectrum_distortion}
\end{figure}

It also provides an experimentally accessible probe of the feedback mechanisms discussed in the previous section. The calculated responsivity and time constant describe the system's response to an infinitesimal change, of which the resonator noise spectrum provides a steady-state manifestation {and means of experimental verification}. In the low power regime, fluctuations in the number density of quasiparticles in the active detector volume map directly into fluctuations in the resonant frequency. At higher drive powers, these fluctuations modulate the resonator impedance at the drive frequency, changing the current flowing through the inductor and altering the total current-dependent inductance of the resonator. The measured spectrum of these fluctuations therefore reflects the full driven response of the detector. Comparison with the low-power spectrum, in which the spectral shape is dominated by intrinsic fluctuations, highlights the impacts of the current nonlinearity. The spectrum of resonant frequency fluctuations {measured in the laboratory using a representative MKID} driven at low, medium, and high generator power is shown in Fig. \ref{fig:spectrum_distortion}.

Independent of any nonlinear feedback effects, because the quasiparticle fluctuations modulate the amplitude and phase of the readout tone as it probes the resonator, increasing the readout tone increases their amplitude above additive system noise sources such as from the LNA. In units of resonant frequency fluctuations, this corresponds to a suppression of system noise sources proportional to the generator power.

As the current-dependent nonlinearity becomes significant, with the resonator biased under positive feedback conditions, a given quasiparticle fluctuation results in an increased fluctuation in the resonant frequency. This results in an increase in the white noise level of the resonator spectrum, seen in the lighter traces in Fig. \ref{fig:spectrum_distortion}.
This increase corresponds to the increase in responsivity (resonant frequency shift per quasiparticle change) due to the nonlinear feedback. It does not require an increase in the scale of the quasiparticle fluctuations themselves.

{Our assertion} that the increased readout current is not increasing the quasiparticle noise itself is supported by the ratio of an extrinsic signal (small peak at {37} Hz from a cryogenic LED) to the noise level. As the nonlinear drive increases, both the signal height and the height of the noise floor increase together. If the dominant noise source is the resonator's own spectrum, this preserves the signal-to-noise ratio. \textbf{If the sensitivity is limited by additive system noise sources such as the LNA, this increases the signal-to-noise ratio.}

The spectra in Fig. \ref{fig:spectrum_distortion} are fitted with a single-pole Lorentzian to extract the dominant time constant, which dictates the roll-off frequency.
Under linear conditions, the rate of decay of perturbation modes in the system is fast, and so the time constant is dominated by the average lifetime of the quasiparticles. Near the critical point, the decay rate slows dramatically (Fig. \ref{fig:time_constant_three_foffsets}), and sets the time constant of the system. This suppresses higher frequency modes, while lower frequency modes receive an enhancement due to the increased responsivity. 
This is seen in the increasing white noise level and decreasing bandwidth as the generator power is increased. In the high power measurement, the spectrum visibly deviates from the single-pole approximation, consistent with $\tau_r$ becoming comparable to and larger than $\tau_{qp}$.

\section{Conclusion}

We have described the nonlinear operational state of a superconducting resonator {detector} in terms of a fixed readout generator frequency and power. In this basis, the readout current is not simply a probe of the resonator state, but also a control parameter. The current-dependent kinetic inductance modifies the resonator's natural frequency, its responsivity to quasiparticle fluctuations, and its bandwidth.

The feedback between the readout tone and the resonance is set by the detuning of the tone from the driven resonance. In the positive feedback regime (when the driven resonant frequency is above the readout tone), perturbations to the resonator state produce enhanced changes in resonant frequency over a reduced bandwidth. In the negative feedback regime (the driven resonance below the tone), the response is suppressed. The same feedback factor controls the response to both readout current changes and to quasiparticle fluctuations. The onset of bifurcation corresponds to unity loop gain of the readout current feedback, with the driven system's response becoming singular.

Operation in the positive feedback regime trades bandwidth and stability margin for low-frequency gain. The resonator's frequency noise spectrum provides a steady-state manifestation of this trade-off{, which we use for laboratory verification of our calculation}. An increasing nonlinear bias leads to an increased white noise level with reduced roll-off frequency, corresponding to an enhanced responsivity and extended relaxation time. This supports consideration of the readout-current nonlinearity as a selectable detector operating state, rather than a limit on the applicable readout power.

\section{Acknowledgments}

The McGill authors acknowledge funding from the Natural Sciences and Engineering Research Council of Canada and the Canada Research Chairs Program.

\bibliography{bibliography}

\appendix

\section{Cavity mode steady-state}\label{app:steady_state_derivation}

As described in \cite{Anferov2020}, the resonator may be understood as the one-port optical cavity,\cite{Eichler2014}\cite{YurkeBuks2006} here coupled symmetrically into a three-port network. The steady state is a balance of the fields, $b(t)$, stored within and entering and exiting the network:

\begin{equation}\label{eq:app:steady_state_a}
    b \left[ j(\omega_g - \omega_0 - K |b|^2) + \dfrac{\omega_0}{2 Q_r}  \right] = -\sqrt{\dfrac{\omega_0}{2 Q_c}} b_1^{in}
\end{equation}

where $b_1^{in}$ is the amplitude applied to the input port of the three-port network, such that $|b_i^{in}|^2  = \dfrac{P_g}{\hbar \omega_g}$ and $|b|^2 = n_{ph}$. The squared modulus of this expression gives the steady-state number of photons in the cavity for a given generator frequency and power:

\begin{equation}
    n_{ph} \hbar \omega_g \dfrac{2 Q_c}{\omega_0}  \left[ (\omega_g - \omega_0 - Kn_{ph})^2 + \left(\dfrac{\omega_0}{2 Q_r}\right)^2  \right] = P_g \quad{.}
\end{equation}

\section{Dynamics}\label{app:dynamics}

{In Sec. \ref{sec:dynamics}, we used the eigenvalues and eigenvectors of the driven resonator to describe the evolution of a perturbation about a steady operating point. In this appendix we derive those results starting from the current-dependent kinetic inductance. We first obtain the equation of motion for the current in the resonator, then use this to obtain the forward scattering parameter and its time derivative, and then finally linearize about the driven steady state.}

{The kinetic inductance fraction is $\alpha = \frac{L_k}{L}$, where $L$ is the total inductance.
Only the kinetic portion of the inductance is sensitive to the current, so the nonlinearity expression (Eq. \ref{eq:LK_Isquared}) for the total driven inductance can be rewritten:}

\begin{equation}
    L(I) \simeq L_0 \left[1 + \alpha \frac{I^2}{I_*^2} \right]
\end{equation}

{For $|\alpha \frac{I^2}{I_*^2}| \ll 1$, $\frac{1}{L(I)} \simeq \frac{1}{L_0} (1 - \alpha \frac{I^2}{I_*^2})$, so the instantaneous driven resonant frequency is:}

\begin{equation}
    \omega_r^2 = \frac{1}{L(I)C} \simeq \omega_0 ^2 \left[ 1 - \alpha \frac{I^2}{I_*^2} \right]
\end{equation}

{Relating this to the standard form of a damped, driven oscillator in the current frame, we obtain:}

\begin{equation}
    \ddot{I} + \kappa_i \dot{I} + I{\omega_0}^2 -\frac{{\omega_0}^2 \alpha}{I_*^2} I^3 =  \sqrt{\frac{\kappa_c\omega_0}{{2} Z_0 Z_r}} \dot{V},
    \label{eqn:secondorder}
\end{equation}
where $I$ is the internal resonator current, $\kappa_i = \omega_0 / {Q_i}$ is the resonator internal loss rate, $\kappa_c = \omega_0 / Q_c$ is the coupling loss rate, $\alpha$ is the kinetic inductance fraction, and $V$ is the feedline voltage at the coupling capacitor node.  By setting {$I(t) = \mathcal{I} e^{j\omega_g t} + \mathcal{I}^* e^{-j\omega_g t}$ and $V(t) = \mathcal{V} e^{j\omega_g t} + \mathcal{V}^* e^{-j\omega_g t}$}
where $\mathcal{I}$ is a slowly varying complex amplitude {and the generator amplitude $\mathcal{V}$ is constant}, and applying the rotating wave approximation, we transform Eq. \ref{eqn:secondorder} into a first order differential equation

\begin{equation} 
\dot{\mathcal{I}} = - \left( \frac{\kappa_i}{2} + j\delta - \frac{jK |\mathcal{I}|^2}{I^2_{ZPF}}  \right) \mathcal{I}+ \frac{\mathcal{V}}{2} {\sqrt{\frac{\kappa_c\omega_0}{2Z_0 Z_r}}},
\label{eqn:firstorder}
\end{equation}

where {$\delta \equiv \omega_g - \omega_0$. ${I_{ZPF}^2} = (\hbar {\omega_0} / 2L)$ is the zero-point current fluctuation amplitude in the resonator mode, and $n_{ph} = |\mathcal{I}|^2/I_{ZPF}^2$ is its average photon occupancy. The Kerr coefficient $K = {-}(3 {\omega_0} \alpha / 2)(I^2_{ZPF}/I^2_*)$ follows from the prefactor of the cubic term that is retained after the rotating wave approximation. \cite{YurkeBuks2006} The expression then becomes: }

\begin{equation} 
{\dot{\mathcal{I}} = - \left( \frac{\kappa_i}{2} + j\delta - jK n_{ph}  \right) \mathcal{I}+ \frac{\mathcal{V}}{2} {\sqrt{\frac{\kappa_c\omega_0}{2 Z_0 Z_r}}}}
\end{equation}

The input and output waves on the feedline are related by
\begin{equation}
    \mathcal{V}_g{^+} + \mathcal{V^-}_{out}  = \mathcal{V^+}_{out} = \mathcal{V}
\end{equation}

where $\mathcal{V}_g{^+}$ is the input wave from the generator.  The circulating current in the resonator can be related to the forward scattered wave and input wave using

\begin{equation}
    \mathcal{I}  = \sqrt{ \frac{2\omega_0}{{Z_r} Z_0 \kappa_c} }(\mathcal{V}^+_{g} -\mathcal{V}^+_{out} )
\end{equation}

{and, since the generator amplitude $\mathcal{V}^+_g$ is constant,}

\begin{equation}
    {\dot{\mathcal{I}} = -\sqrt{ \frac{2\omega_0}{{Z_r} Z_0 \kappa_c}} \dot{\mathcal{V}}^+_{out} \quad{.}}
\end{equation}

Substituting {these expressions} into eqn.~\ref{eqn:firstorder},

\begin{align}
    \dot{\mathcal{V}}^+_{out} &= (\frac{\kappa_i}{2} {+} j\delta - jK'|\mathcal{V}{^+}_g - \mathcal{V}^+_{out}|^2)(\mathcal{V}{^+}_g - \mathcal{V}^+_{out}) {-} \frac{\kappa_c}{2}\mathcal{V}^+_{out} \nonumber \\
    &= -(\frac{\kappa}{2} {+} j\delta - jK'|\mathcal{V}{^+}_g - \mathcal{V}^+_{out}|^2)\mathcal{V}^+_{out} + \dots \nonumber \\ %
    & \quad\quad (\frac{\kappa_i}{2} {+} j\delta - jK'|\mathcal{V}{^+}_g - \mathcal{V}^+_{out}|^2)\mathcal{V}{^+}_g, \label{eqn:Vdot}
\end{align}

where $\kappa = \kappa_i + \kappa_c$ is the total loss rate and {$K' = 2K\omega_0 / (Z_r Z_0 \kappa_c I^2_{ZPF})$}.

Setting the time derivative to zero gives the {implicit} steady state solution for the forward scattering parameter
\begin{align}
    S_{21} = \frac{\mathcal{V}^+_{out}}{\mathcal{V}{^+}_g} 
    &= \frac{\kappa_i/2 {+} j\delta - jK'|\mathcal{V}{^+}_g - \mathcal{V}^+_{out}|^2}{\kappa/2 {+} j\delta - jK'|\mathcal{V}{^+}_g - \mathcal{V}^+_{out}|^2} \nonumber \\
    &= 1 - \frac{\kappa_c }{\kappa {+} 2j\delta - 2jK'|\mathcal{V}{^+}_g - \mathcal{V}^+_{out}|^2}.
\end{align}

{As expected for a nonlinear Kerr cavity, the driven resonant frequency depends on the circulating field, and the circulating field depends on the generator detuning from the driven resonant frequency. Reverting to the notation used in the main body of the text, we obtain:}

\begin{align}
        {S_{21}} &= {1 - \dfrac{\kappa_c}{\kappa + 2j(\delta - Kn_{ph})}  } \label{eq:s21_reduced_variables} \\
        &= {1 - \dfrac{\omega_0/Q_c}{\omega_0/Q_r \left[1 + 2jQ_r \frac{\omega_g - \omega_r}{\omega_0} \right]}} \\
        &\simeq { 1 - \dfrac{Q_r}{Q_c}\dfrac{1}{1 + 2jQ_r x} }
\end{align}

{as  in Eq. \ref{eq:s21_nonlinear}.}

{To examine the measurable dynamics of the system, f}rom eqn.~\ref{eqn:Vdot}, we express the time evolution of the scattering parameter as

\begin{align}
    \dot{S}_{21} = \left( \frac{\kappa}{2} {+} j\delta -j {K'|\mathcal{V}_g^+|^2}|1 - S_{21}|^2\right)(1 - S_{21}) - \frac{\kappa_c}{2}.
\end{align}

Setting $U = 1 - S_{21}$ and {$\Psi \equiv K' |\mathcal{V}_g^+|^2$}
, the time evolution equation is rewritten as 

\begin{align}\label{eq:eom_U}
    \dot{U} &= \left( -\frac{\kappa}{2} {-}j\delta + j\Psi UU^*\right) U + \frac{\kappa_c}{2}. \nonumber\\
    &=F(U,U^*)
\end{align}

\begin{figure}[htbp]
    \centering
    \includegraphics[width=\linewidth]{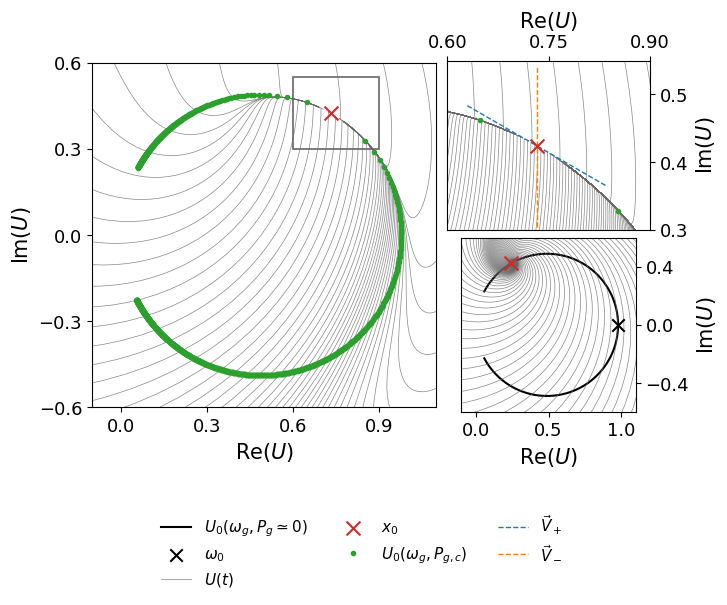}
    \caption{Dynamics of the resonator in the complex ($U = 1 - S_{21}$) plane. Grey curves show numerical trajectories $U(t)$ obtained by integrating Eq. \ref{eq:eom_U} from a circle of initial conditions centred on the origin, and the red $\times$ indicates the location of the generator in the resonator frame. The lower right panel shows the low power case, where the generator is located at $x_{0,c}$ but $P_g \simeq 0$. In the main panel, the system is biased at the critical point ($x_{0,c}, P_{g,c}$); the green points show the corresponding nonlinear steady state solutions during a finite-resolution frequency sweep at the same generator power. The upper right panel shows an enlarged view of the boxed region, with the directions of the eigenvectors overlaid as dashed lines. At the critical point, the $U$ trajectories align with these vectors, either parallel the imaginary axis (for $\vec{V}_-$) or tangent to the resonance circle (for $\vec{V}_+$), although since the corresponding eigenvalue $\lambda_+ = 0$, the latter are not visible. }
    \label{fig:U_solutions}
\end{figure}

Fig. \ref{fig:U_solutions} shows numerical solutions $U(t)$ to Eq. \ref{eq:eom_U} for a set of starting coordinates, with the system at the critical point 
and at low generator power. 
The directions of the two eigenvectors, derived below, are also indicated at the critical point.

We consider a small perturbation around the steady state solution, $U = U_0 + \epsilon$.  The derivatives
\begin{align}
    \frac{\partial F}{\partial U} &= -\frac{\kappa}{2} {-} j\delta + 2j\Psi UU^* \nonumber \\
    \frac{\partial F}{\partial U^*} &= j\Psi U^2
\end{align}

provide the linearized equation {for the evolution of the perturbation:}

\begin{equation}
\dot{\epsilon} \simeq \left. \frac{\partial F}{\partial U} \right|_{0} \epsilon + \left. \frac{\partial F}{\partial U^*} \right|_{0} \epsilon^*.
\end{equation}

In terms of the real and imaginary parts{, $U_0 = U_x + j U_y$ and} $\epsilon = \epsilon_x + j\epsilon_y$

\begin{equation}\label{eqn:lintdep}
    \begin{bmatrix}
    \dot{\epsilon}_x \\
    \dot{\epsilon}_y
    \end{bmatrix}
    = %
    M
    \begin{bmatrix}
    {\epsilon}_x \\
    {\epsilon}_y
    \end{bmatrix},
\end{equation}

where
\begin{align}
    M &= -\frac{\kappa}{2}
    \begin{bmatrix}
        1 & 0 \\
        0 & 1
    \end{bmatrix}
    + ({-}\delta + 2\Psi |U_0|^2)
    \begin{bmatrix}
        0 & -1 \\
        1 & 0
    \end{bmatrix} \dots \\
    & \quad+ \Psi
    \begin{bmatrix}
    -2 U_x U_y & U_x^2 - U_y^2 \\
    U_x^2 - U_y^2 & 2U_x U_y
    \end{bmatrix}.
\end{align}

The solutions to \ref{eqn:lintdep} are
\begin{equation}
    \epsilon(t) = {c_+} e^{\lambda_+ t} \vec{V}_+ + {c_-} e^{\lambda_- t} \vec{V}_-,
\end{equation}

where {the coefficients $c_{\pm}$ are the arbitrary mode amplitudes}, the eigenvalues of $M$ are
\begin{equation}\label{eq:eigenvalues_U}
    {\lambda_{\pm}} = -\frac{\kappa}{2} \pm \sqrt{-(\delta {-} \Psi |U_0|^2) (\delta {-} 3\Psi|U_0|^2)},
\end{equation}

\noindent and ${\vec{V}_{\pm}}$ are the corresponding eigenvectors.

{Expanding the condensed variables using $\Psi |U_0|^2 = K n_{ph}$, we obtain the expression used in Sec. \ref{sec:dynamics}:}

\begin{equation}
    {\lambda_{\pm} = - \dfrac{\omega_0}{2 Q_r} \pm \sqrt{(Kn_{ph})^2 - (\omega_0 - \omega_g + 2 Kn_{ph})^2}}
\end{equation}

{We now examine the evolution of a perturbation about this critical point.} 
Using Eq. \ref{eq:s21_reduced_variables}, we obtain the coordinates of the critical point  $U_c = (U_{x,c} , U_{y,c}) = \frac{\kappa_c}{\kappa} ( \frac{3}{4}, -\frac{K}{|K|} \frac{\sqrt{3}}{4} )$

{The tangent vector at this point is }
\begin{equation}
    \vec{t}_c \propto \begin{bmatrix} -2U_{x,c}U_{y,c} \\ U_{x,c}^2 - U_{y,c}^2 \end{bmatrix} \propto \begin{bmatrix}
        \sqrt{3} \\ \frac{K}{|K|} 
    \end{bmatrix}
\end{equation}

\noindent{and the Jacobian $M$ is}
\begin{equation}
    M_c = \\
    \begin{bmatrix}
        0 & 0 \\
        \frac{K}{|K|} \frac{\kappa}{\sqrt{3}} & -\kappa \\
    \end{bmatrix}
\end{equation}

{From this we obtain the eigenvector corresponding to the slow eigenvalue $\lambda_+ = 0$ as:}

\begin{equation}
    \vec{V}_+ \propto
    \begin{bmatrix}
         \sqrt{3} \\ 
         \frac{K}{|K|}
    \end{bmatrix}
\end{equation}

\noindent{which is identical to the tangent vector, showing that this eigenvector is parallel to the tangent direction. A perturbation tangent to the resonance circle at the critical point (that is, in the resonator's phase direction) therefore experiences no restoring force, and does not decay.}

{Conversely, the other eigenvector must satisfy: }

\begin{equation}
    M_c \vec{V}_- = -\kappa \vec{V_-}
\end{equation}

\noindent for which 

\begin{equation}
    \vec{V}_- = \begin{bmatrix}
        0 \\ 1
    \end{bmatrix}
\end{equation}

\noindent is a solution.

{A perturbation along this direction evolves as $\epsilon(t) \propto e^{-\kappa t} \begin{bmatrix}
    0\\1
\end{bmatrix}$. This mode has a decay time constant $\tau_- =  1/\kappa = Q_r / \omega_0$, which is half the ordinary low-power decay time. Its direction is along $U_y$, aligned with the imaginary axis but not strictly perpendicular to the resonance circle, as seen in Fig. \ref{fig:U_solutions}.}

\end{document}